\documentstyle[preprint,tighten,eqsecnum,aps]{revtex}
\begin{document}
\draft
\title{Gravitational Waves from an Axi--symmetric Source \\ in the
Nonsymmetric Gravitational Theory}
\author{ N. J. Cornish, J. W. Moffat and D. C. Tatarski}
\address{Department of Physics, University of Toronto \\ Toronto,
Ontario M5S 1A7, Canada}
\maketitle
\begin{abstract}
We examine gravitational waves in an isolated axi--symmetric
reflexion symmetric NGT system. The structure of the vacuum field
equations is analyzed and the exact solutions for the field
variables in the metric tensor are found in the form of expansions
in powers of a radial coordinate. We find that in the NGT axially
symmetric case the mass of the system remains constant only if the
system is static (as it necessarily is in the case of {\em spherical}
symmetry). If the system radiates, then the mass decreases monotonically
and the energy flux associated with waves is positive.
\end{abstract}

\narrowtext

\section{Introduction}

The present work examines gravitational radiation in the
Nonsymmetric Gravitational Theory (NGT) (for a recent detailed
review see \cite{Moff91}). We probe the NGT asymptotic behaviour in
the wave zone using an {\em exact} solution. A complementary analysis
based on a DeWitt style background field expansion has already been
published, announcing the main result of this current work \cite{pla}.

The motivation for this work is twofold. Firstly, the
nature of gravitational radiation is an important physical
question which must be addressed in any candidate gravitational theory.
Secondly, the literature already contains several incorrect treatments
of gravitational radiation in NGT \cite{Krisher,DDM1,DDM2}, and since more
incorrect treatments are currently being published \cite{Dam},
it is important that the record be set straight.
The reasons why erroneous results were arrived at by these other authors are
explained in \cite{CorMoff}.

In General Relativity (GR) gravitational radiation from bounded
sources has been studied not only through the linearized theory but
also with the use of exact solutions. The latter was done for the
general case of a bounded source in asymptotically flat spacetime
\cite{Sachs}. It was found that confining the arguments to the
axially symmetric case did not cause any essential loss of
generality. Since even the relevant GR calculations are very
tedious and the level of technical difficulty in the case of NGT
increases considerably, we limit ourselves to the axi--symmetric
case. The GR gravitational waves from isolated axially symmetric
reflexion symmetric systems were studied in detail in \cite{BBM}.
Since our treatment of the axi--symmetric NGT case is rather
parallel, familiarity with this analysis is strongly recommended.

Since NGT was introduced \cite{Moff79-1} there have been few
analytic solutions of the field equations published. The exact
solutions known to date include the spherically symmetric vacuum
case \cite{Moff79-2}, the spherically symmetric interior case
\cite{Sav89,Sav90} and Bianchi type I cosmological solutions with
and without matter \cite{Kunst79,Kunst80}.
This, at least in part, follows from the fact that deriving NGT
field equations relevant for particular cases of interest is not as
technically simple as may be suggested by its superficial
similarity to the corresponding GR situations. Firstly, since the
underlying geometry is non-Riemannian, neither the fundamental
metric tensor \(g_{\mu \nu}\) nor the affine connection is
symmetric. This does not constitute a serious problem for the
choice of the form of \(g_{\mu \nu}\), since we can always assume
that its nonsymmetric part takes on the isometries of the symmetric
part, which in turn has a well defined GR limit. On the other hand,
calculating the connection coefficients proves to be a tedious and
time consuming exercise, independent of the method chosen.
Secondly, the resultant formulae for the nonsymmetric connection
coefficients are extremely complicated for all but the simplest
forms of the metric, thus becoming unwieldy to use in the
derivation of still more complicated field equations.

The NGT quantities presented in this paper were derived with the
use of symbolic algebraic computation procedures. To this end, we
have used the symbolic computation system {\em Maple}.

In Section \ref{fielde}, we briefly summarize the necessary
fundamentals of NGT. Section \ref{coord} deals with the coordinate
system and generalization of the GR metric to the NGT case. Then in
Section \ref{res}, we expand the metric in negative powers of a
suitably chosen radial coordinate and analyze the field equations.
The closing section contains our conclusions.

Throughout this paper we use units in which \(G=c=1\).

\section{NGT Vacuum Field Equations} \label{fielde}

The NGT Lagrangian without sources takes the form:
\begin{equation} \label{lagr}
{\cal L} =  \sqrt{-g}g^{\mu \nu} R_{\mu \nu}(W),
\end{equation}
with \(g\) the determinant of \(g_{\mu \nu}\). The NGT Ricci tensor
is defined as:
\begin{eqnarray} \label{ricciw}
R_{\mu \nu}(W) &=& W^{\beta}_{\mu \nu , \beta}- \frac{1}{2}
(W^{\beta}_{\mu \beta , \nu}+W^{\beta}_{\nu \beta , \mu}) \nonumber\\
&& \hspace{.2in} -
W^{\beta}_{\alpha \nu}W^{\alpha}_{\mu \beta}+W^{\beta}_{\alpha
\beta}W^{\alpha}_{\mu \nu},
\end{eqnarray}
and \(W^{\lambda}_{\mu \nu}\) is an unconstrained nonsymmetric
connection :
\begin{equation} \label{connw}
W^{\lambda}_{\mu \nu}=W^{\lambda}_{(\mu \nu)}+W^{\lambda}_{[\mu
\nu]}.
\end{equation}
(Throughout this paper parentheses and square brackets enclosing
indices stand for symmetrization and antisymmetrization,
respectively.)
The contravariant nonsymmetric tensor \(g^{\mu \nu}\) is defined in
terms of the equation:
\begin{equation} \label{inverse}
g^{\mu \nu} g_{\sigma \nu}=g^{\nu \mu} g_{\nu
\sigma}=\delta^{\mu}_{\sigma}.
\end{equation}
If we define the torsion vector as:
\begin{equation}
W_{\mu} \equiv W^{\nu}_{[\mu \nu]} = \frac{1}{2} \left(
W^{\nu}_{\mu \nu}- W^{\nu}_{\nu \mu} \right),
\end{equation}
then the connection:
\begin{equation} \label{conng}
\Gamma^{\lambda}_{\mu \nu} = W^{\lambda}_{\mu \nu} + \frac{2}{3}
\delta^{\lambda}_{\mu} W_{\nu}
\end{equation}
is torsion free:
\begin{equation} \label{gtors}
\Gamma_{\mu} \equiv \Gamma^{\alpha}_{[\mu \alpha]} = 0.
\end{equation}
Defining now:
\begin{eqnarray} \label{riccig}
R_{\mu \nu}(\Gamma) &=& \Gamma^{\beta}_{\mu\nu,\beta} -
\frac{1}{2}(\Gamma^{\beta}_{(\mu\beta),\nu} + \Gamma^{\beta}_{(\nu
\beta) , \mu}) \nonumber\\
&& \hspace{.2in} -\Gamma^{\beta}_{\alpha\nu}\Gamma^{\alpha}_{\mu
\beta}+\Gamma^{\beta}_{(\alpha\beta)}\Gamma^{\alpha}_{\mu\nu},
\end{eqnarray}
we can write:
\begin{equation} \label{ricciw=g}
R_{\mu \nu}(W) = R_{\mu \nu}(\Gamma) + \frac{2}{3} W_{[\mu ,\nu]},
\end{equation}
where \(W_{[\mu,\nu]}=\frac{1}{2}(W_{\mu,\nu}-W_{\nu,\mu})\).
Finally, the NGT vacuum field equations can be expressed as:
\begin{equation} \label{fensgamma}
g_{\mu\nu,\sigma} - g_{\rho\nu} {\Gamma}^{\rho}_{\mu\sigma} -
g_{\mu\rho} {\Gamma}^{\rho}_{\sigma\nu} = 0 ,
\end{equation}
\begin{equation} \label{fensdiver}
{(\sqrt{-g}g^{[\mu \nu]})}_{ , \nu} = 0 ,
\end{equation}
\begin{equation} \label{fensricci}
R_{\mu \nu}(\Gamma) = \frac{2}{3} W_{[\nu , \mu]}.
\end{equation}
For the purpose of the analysis of Section \ref{res}, it is
convenient to decompose \(R_{\mu\nu}\) into standard symmetric and
antisymmetric parts: \(R_{(\mu\nu)}\), \(R_{[\mu\nu]}\), and then
rewrite the field equation (\ref{fensricci}) in the following form:
\begin{equation} \label{sym}
R_{(\mu \nu)}(\Gamma) = 0,
\end{equation}
\begin{equation} \label{asym}
R_{[\mu \nu , \rho]}(\Gamma) = 0,
\end{equation}
where we used equations (\ref{conng}), (\ref{gtors}) and the
notation:
\begin{equation}
R_{[\mu\nu,\rho]} = {R_{[\mu\nu]}}_{,\rho} + {R_{[\nu\rho]}}_{,\mu}
+ {R_{[\rho\mu]}}_{,\nu} .
\end{equation}

\section{The Metric} \label{coord}

Similarly to GR, the simplest NGT field due to a bounded source
would be spherically symmetric. However, the NGT equivalent of
Birkhoff's theorem (see e.g. \cite{Moff91}) shows that a
spherically symmetric gravitational field in an empty space must be
static. Hence, no gravitational radiation escapes into empty space
from a pulsating spherically symmetric source.

Following \cite{BBM} we consider the next simplest case: the field
which was initially static and spherically symmetric and eventually
becomes such, but undergoes an intermediate non--spherical wave
emitting period. Also, spacetime is assumed to be axially symmetric
and reflexion--symmetric at all times. Because of the
complexity of the field equations, we are forced to use the method
of expansion to examine the problem. This approach, namely
expanding in negative powers of a radial coordinate, was also used
in the GR analysis \cite{BBM} and naturally suits a wave
problem.

Due to the physical picture sketched above and to the fact that we
are interested in the asymptotic behaviour of the field at null
infinity, ${\cal I}$, (in an arbitrary direction from our isolated source)
polar coordinates \(x^{0}=u, {\bf x} = (r,\theta,\phi)\) are the natural
choice. The ``retarded time'' \(u=t-r\) has the property that the
hypersurfaces \(u=\mbox{constant}\) are light--like. Detailed
discussion of the coordinate systems permissible for investigation
of outgoing gravitational waves from isolated systems can be found
in \cite{Sachs,BBM}.

The covariant GR metric tensor corresponding to the situation
described above is:
\begin{equation} \label{grmetric}
g_{\mu\nu}\! = \!\left( \begin{array}{cccc}
Vr^{-1}e^{2\beta}-U^{2}r^{2}e^{2\gamma} & e^{2\beta} &
Ur^{2}e^{2\gamma} & 0 \\
e^{2\beta} & 0 & 0 & 0 \\
Ur^{2}e^{2\gamma} & 0 & -r^{2}e^{2\gamma} & 0 \\
0 & 0 & 0 & -r^{2}e^{-2\gamma}\sin^{2}\theta \end{array} \right)
\end{equation}
with \(U,V,\beta,\gamma\) being functions of \(u,r\) and
\(\theta\) was first given in \cite{Bondi60}.

For any metric in polar coordinates, form conditions must be imposed
in the neighbourhood of the polar axis, \(\sin\theta=0\), to ensure
regularity. In the case under consideration we have that, as
\(\sin\theta\rightarrow0\), $ V,\beta,U/\sin\theta,\gamma/\sin^{2}\theta$
each is a function of \(\cos\theta\) regular at \( \cos\theta = \pm
1 \).

In order to find the NGT generalization of the metric tensor
(\ref{grmetric}) we require that the symmetric part of the NGT
metric tensor be formally the same as the GR metric tensor. We
then impose the spacetime symmetries of the symmetric metric onto
the antisymmetric sector. This is achieved by enforcing
$\pounds_{ {\vec{\xi}_{(i)}} }g_{[\mu\nu]}=0$, where the Killing vector
field, $\vec{\xi}_{(i)}$, is obtained from $\pounds_{ {\vec{\xi}_{(i)}} }
g_{(\mu\nu)}=0$. The solution to this equation for the metric (\ref{grmetric})
yields the single Killing vector field $\vec{\xi}_{(1)}=\xi^{3}_{(1)}
\partial_{\phi}=\sin^{2}\theta \partial_{\phi}$. Imposing
$\pounds_{{\vec{\xi}_{(1)}} }g_{[\mu\nu]}=0$ yields:
\begin{equation}
\xi^{3}_{(1)}\partial_{\phi} g_{[\mu\nu]} +g_{[\mu 3]}\partial_{\nu}
\xi^{3}_{(1)} +g_{[3\nu]}\partial_{\mu}\xi^{3}_{(1)}=0 \; .
\end{equation}
This equation gives $\partial_{\phi}g_{[\mu\nu]}=0$, but does not exclude
any antisymmetric components. This is markedly different from
the static spherically symmetric case where the above procedure excludes
four of the six antisymmetric components. Without further
simplification, the NGT calculation would involve ten independent
functions and ten independent non-linear differential equations.
This would constitute a huge increase in complexity from the system of
four equations and functions found in the GR case.

To make the problem tractable, we need to determine which antisymmetric
functions can be set to zero.
To accomplish this we note that the imposition of axi-symmetry splits
the antisymmetric field equations (\ref{fensdiver}), (\ref{asym})
into two sets of three independent equations. (This can be seen directly
from the block-diagonal form of the GR metric). The first set explicitly
involves the three skew functions $g^{[01]}\, , g^{[02]}\, , g^{[12]}$:
\begin{eqnarray}
\left(\sqrt{-g}g^{[\mu\nu]}\right)_{,\nu}&=&0 \hspace{1in}
(\mu=0,1,2) \; ,  \nonumber \\ \nonumber \\
R_{[01,2]}&=&0 \; .
\end{eqnarray}
These four equations are not independent due to the one identity:
\begin{equation}
\left(\sqrt{-g}g^{[\mu\nu]}\right)_{,\nu,\mu}=0 \hspace{1in} (\mu,\nu=0,1,2)
\; .
\end{equation}
The second set of four equations explicitly involves the three skew
functions $g^{[30]}\, ,g^{[31]}\, ,g^{[32]}$:
\begin{eqnarray}
\left(\sqrt{-g}g^{[3 \nu]}\right)_{,\nu}&=&0 \; , \nonumber \\
\nonumber \\
R_{[3 \mu ,\nu]}&=&0 \; .
\end{eqnarray}
These four equations are also not independent due to the one
identity
\begin{equation}
\epsilon^{3 \mu\nu\rho}R_{[3 \mu ,\nu],\rho}=0 \; .
\end{equation}
We may now choose to work with either set of three equations and three
functions, noting that eliminating one set of three functions simultaneously
eliminates the three corresponding equations. We choose to work with the
first set of functions and equations as they reproduce the usual static
spherically symmetric solution when the $u$ and $\theta$ dependence is
suppressed. The components that we discard correspond to the magnetic
monopole-like solutions of NGT. These components can be discarded without
loss of generality as an extension of the background field analysis of
Ref.\cite{pla} to include all six skew functions does not alter our
conclusions \cite{endnote}.

\widetext

In view of the above, the NGT generalization of the metric tensor
(\ref{grmetric}) is:
\begin{equation} \label{genmetric}
g_{\mu\nu} = \left( \begin{array}{cccc}
Vr^{-1}e^{2\beta}-U^{2}r^{2}e^{2\gamma} & e^{2\beta} + \omega &
Ur^{2}e^{2\gamma} + \lambda & 0 \\
e^{2\beta} - \omega & 0 & \sigma & 0 \\
Ur^{2}e^{2\gamma} - \lambda & -\sigma & -r^{2}e^{2\gamma} & 0 \\
0 & 0 & 0 & -r^{2}e^{-2\gamma}\sin^{2}\theta \end{array} \right),
\end{equation}
where \(\omega ,\lambda\) and $\sigma$ are functions of \(u,r\) and
\(\theta\). The contravariant metric tensor is given by:
\begin{equation} \label{contrmetric}
g^{\mu\nu} = \left( \begin{array}{cccc}
-\sigma^{2}e^{-2\gamma}A & g^{01}
& g^{02}  & 0 \\
g^{10} & (re^{2\beta}V-\lambda^{2}e^{-2\gamma})A
& g^{12} & 0 \\
g^{20} &
g^{21} & -(e^{4\beta}-\omega^{2})A e^{-2\gamma} & 0 \\
0 & 0 & 0 & -r^{-2}e^{2\gamma}\sin^{-2}\theta \end{array} \right),
\end{equation}
where
\begin{eqnarray}
A&=&{r^{2}\sin^{2}\theta \over g} , \nonumber \\
g&=&r^{2}\sin^{2}\theta [ -r^{2}(e^{4\beta}-(\omega-\sigma U)^{2})
+\sigma e^{2\beta -2\gamma}(2\lambda -\sigma {V \over r}) ] , \nonumber \\
g^{01}&=&[r^{2}(\omega-\sigma U-e^{2\beta})+\sigma\lambda e^{-2\gamma}]
A , \nonumber \\
g^{02}&=&(e^{2\beta}-\omega)\sigma e^{-2\gamma}A ,\nonumber \\
g^{12}&=& [ Ur^{2}(\sigma-\omega-e^{2\beta})+e^{2\beta-2\gamma}(\lambda
-\sigma Vr^{-1})+\lambda\omega e^{-2\gamma} ]A , \nonumber \\
g^{\mu\nu}&=&g^{\nu\mu}[(\omega, \sigma, \lambda) \rightarrow
(-\omega, -\sigma, -\lambda)]. \nonumber
\end{eqnarray}
\narrowtext

\section{The Field Equations} \label{res}

The analysis determining the form of the functions
\(U,V,\beta,\gamma,\omega,\lambda,\sigma\) in our case is a natural
extension of that given in detail by Bondi {\it et. al.}\cite{BBM} for finding
the forms of \(U,V,\beta,\gamma\). The requirement that the field contain
only outgoing radiation at large distances from the source gives the
form of \(\gamma\):
\[
\gamma=\frac{f(u,\theta)}{r}+\frac{g(u,\theta)}{r^{3}}+...
\]
Demanding that the solution have the correct static limit (or equilibrium
configuration) leads to the following forms for $U,\, V,\, \beta$ and
$\gamma$ (unless otherwise stated, all coefficients in the general expansions
are functions of both $u$ and $\theta$):
\begin{eqnarray}
U&=&{U_{1} \over r}+{U_{2} \over r^2} +\dots \, , \\
V&=& r-2M+{V_{1} \over r} +\dots \, , \\
\beta&=& {B_{1} \over r}+{B_{2} \over r^2} +\dots \, , \\
\gamma&=& {c \over r}+{C-{1\over 6}c^3 \over r^3} +\dots \, ,
\end{eqnarray}

The skew functions, $\omega,\;\lambda$ and $\sigma$, are constrained
by the requirement that the spacetime is asymptotically Lorentzian and
admits inhomogeneous orthochronous Lorentz transformations \cite{Sachs}.
This requirement demands that $ g_{[\mu\nu]}g^{[\mu\nu]}
\rightarrow 0$ as $r \rightarrow \infty$. In our present coordinates this
condition is satisfied if $\omega,\, \lambda$ and $\sigma$ have the
following forms:
\begin{eqnarray}
\omega&=& {W_{1} \over r}+{W_{2} \over r^2}+\dots \, ,\\
\lambda&=& L_{0}+{L_{1} \over r}+ \dots \, ,\\
\sigma&=& S_{0}+{S_{1} \over r}+ \dots \, .
\end{eqnarray}
The functions $M,\, c,\, C,\, L_{0},$ and $W_{2}$ are all functions of
integration. Bondi {\it et. al.} refer to $c$ as the ``news function'' as it
controls the form of the gravitational radiation in the symmetric
sector. In an analogous way, $L_{0}$ is the ``news function'' for the
antisymmetric sector. Consistent with these identifications, we shall see that
the solution reduces to the static, non-radiative case when both $c$ and
$L_{0}$ are set to zero. The static limit tells us that only $M$ and
$W_{2}$ can be non-zero when the system passes through its equilibrium
position and these coefficients will be identified as the mass and NGT charge
of the body, respectively.

   We begin our analysis of the field equations by considering the simplest
set of field equations -- the skew divergence equations $(\sqrt{-g}
g^{[\mu\nu]})_{,\nu}=0$. The $\theta$ component of this set becomes:
\begin{equation}
0=S_{0\; ,u} - {2 (S_{0}c)_{,u}-S_{1\; ,u} \over r} +\dots \, ,
\end{equation}
and since $\sigma$ must equal zero when passing through equilibrium, $S_{0}=
S_{1}=0$ always. Inserting this information into the $u$ component directly
returns $W_{1}=0$ from the lowest order term. The remaining, $r$, component
yields:
\begin{equation}
0= L_{0}\cot\theta+L_{0\; ,\theta}-W_{2\; ,u} +\dots \, .
\end{equation}

At this stage of the calculation, it is not profitable to continue to work
with the skew divergence equations, as the next orders also contain unknown
coefficients from the symmetric functions. Somewhat surprisingly, however,
we are already in a position to calculate the NGT charge of the body, and to
prove that it is conserved. The NGT charge, $l^2$, is defined by the Gaussian
surface integral
\begin{equation}
l^2\equiv {1 \over 4\pi} \int (\sqrt{-g} g^{[0\nu]})_{,\nu} d^3 x=
{1 \over 2} \int_{0}^{\pi} W_{2} \sin\theta\, d\theta = <W_{2}> , \label{ngtc}
\end{equation}
where the brackets $<>$ denote the angular average. The charge is conserved
since
\begin{equation} \label{ngtcc}
l^2_{,u}={1 \over 2}\int_{0}^{\pi} W_{2\; ,u} \sin\theta\, d\theta=
{1 \over 2}\int_{0}^{\pi} (L_{0}\sin\theta)_{,\theta}\, d\theta=0 \; .
\end{equation}
This follows from the fact that $L_{0}$ must be regular on the polar axis.

We now turn our attention to the set of field equations $R_{(\mu\nu)}=0$.
The affine connections that we require to construct these generalized Ricci
tensor components are obtained by solving the system
of 64 equations (\ref{fensgamma}). The closed form expressions for the
non-zero connection components are extremely lengthy.
For brevity, we only display the connections
expanded in inverse powers of $r$, and only to the order required
for our present analysis. The list of expanded non-zero components
is given in the appendix.

Rather than provide an exhaustive list of the $R_{(\mu\nu)}=0$ equations,
we shall simply exhibit the components to the order that we require
to obtain the solution. We begin with the
$(rr)$ component which demands:
\begin{equation}
0={2B_{1} \over r^3}+{c^2+4B_{2} \over r^4}+{6B_{3} \over r^5}+\dots \, ,
\end{equation}
this gives $B_{1}=B_{3}=0$ and $B_{2}=-{c^2/4}$. The $(r \theta)$ component
then gives
\begin{equation}
0={U_{1} \over r} +{U_{2}+U_{1}c+c_{,\theta}+2c \cot\theta \over r^2}+\dots
\, ,
\end{equation}
which yields $U_{1}=0$ and $U_{2}=-(c_{,\theta}+2c \cot\theta)$. Inserting
these expressions into the $(uu)$ component results in the important condition
\begin{equation}
\label{massderiv}
M_{,u}= -{c_{,u}}^{2} + \frac{1}{2} {(c_{,\theta\theta}
+3c_{,\theta}\cot\theta -2c)}_{,u} \;   .
\end{equation}
We can now use the $(ur)$ components of
$R_{(\mu\nu)}=0$ at next order to solve for $V_1$, but first we must choose
how we wish to write the function of integration contained in $U_3$.
Following Bondi {\it et. al.} we shall write $U_{3}$ as
\begin{equation}
U_{3}=2N + 3cc_{,\theta} +4c^{2}\cot\theta \; ,
\end{equation}
where $N$ is the additional function of integration (the reason that $U_{3}$
is written in this way rather than as $U_{3}=\widetilde{N}$, and that the
second function of integration in $\gamma$ is written as $C-{1\over 6}c^3$
rather
than as $\widetilde{C}$, is that we can then identify $N$ and $C$ as the
dipole and quadrupole moments of the source, respectively). Now the $(ur)$
equation at next order:
\begin{eqnarray}
0&=& \sin^{2}\theta\left(11c^2 -5(c_{,\theta})^2+4c^{2}c_{,u}
-6cc_{,\theta\theta}\right) \nonumber \\
&&+U_{3}\sin{\theta}\cos{\theta} -8c^2-
19cc_{,\theta}\cos{\theta}\sin{\theta} \nonumber\\
&&+\sin^{2}\theta\left(U_{3\, ,\theta}-2L_{0}^2+2V_{1}\right) \; ,
\end{eqnarray}
can be solved to give
\begin{eqnarray}
V_{1}&=&-N_{,\theta} -N\cot\theta +{c_{,\theta}}^{2} +
4cc_{,\theta}\cot\theta \nonumber\\
 &&+\frac{1}{2}c^{2}(1+8\cot^{2}\theta)+L_{0}^2 \; .
\end{eqnarray}
It is interesting to note that $V_{1}$ contains the first explicit difference
between the symmetric functions found for GR and those found for NGT.
Substituting the above results into the $(\theta\theta)$ and $(u \theta)$
components of $R_{(\mu\nu)}=0$ produces the following auxiliary conditions
on the multipole moments $C,\, N$:
\begin{eqnarray}
4C_{,u}&=& 2c^{2}c_{,u}+2cM+N\cot\theta-N_{,\theta}-L_{0}^2 \; ,\\
-3N_{,u}&=& M_{,\theta} +3cc_{,u\theta} +4cc_{,u}\cot\theta
+c_{,u}c_{,\theta} \;  .
\end{eqnarray}
We see that the quadrupole moment of a source will decrease more rapidly in
NGT than in GR due to the $L_{0}^2$ term.

For completeness, we shall now return to the antisymmetric sector where the
skew divergence equation can now be used to obtain the additional relations:
\begin{eqnarray}
W_{3}&=& 2S_{2\, ,\theta}+S_{2}\cot{\theta} \; , \\
S_{2\, , u}&=& L_{1}-2cL_{0} \; .
\end{eqnarray}
The only remaining antisymmetric field equation (\ref{asym}):
\begin{equation} \label{anti}
R_{[01,2]}=R_{[01],2}+R_{[12],0}+R_{[20],1}=0 ,
\end{equation}
gives to lowest order:
\begin{equation}
\left(W_{2\, ,\theta}+L_{1}+S_{2\, ,u}\right)_{,u}=0 \; .
\end{equation}
This equation yields the additional relation:
\begin{equation}
W_{2\, ,\theta}=2cL_{0}-2L_{1} \; .
\end{equation}

\subsection*{Analysis of the solution}

To demonstrate the physical interpretation of \(M\), we consider the
static limit. We can scale the function \(c\) for either one of the
static periods to be \(c=0\) (forsaking the \(\theta\) dependence
of \(c\) limits us here to a static spherically symmetric system).
We now remove the terms containing the functions \(N\) and \(C\) since
they correspond to multipole moments.
Since there is no radiation during the static period, \(N=C=0\).
The metric (\ref{genmetric}) tends now to its static spherically
symmetric limit:
\begin{eqnarray}
\label{expstmetric00}
g_{00}&=&1-\frac{2M_{s}}{r} +\frac{l_{s}^{4}}{r^{4}} -
\frac{2M_{s}l_{s}^{4}}{r^{5}} , \\
\label{expstmetric01s}
g_{(01)}&=&1+\frac{l_{s}^{4}}{2r^{4}} , \\
\label{expstmetric01as}
g_{[01]}&=&\frac{l_{s}^{2}}{r^{2}} , \\
\label{expstmetric02}
g_{(02)}&=&g_{[02]}=0 ,  \\
\label{expstmetric22}
g_{22}&=& -r^{2} , \\
\label{expstmetric33}
g_{33}&=& -r^{2} \sin^{2}\theta  ,
\end{eqnarray}
where by \(M_{s},\, l^{2}_{s}\) we denote the static limit of \(M\) and
$l^{2}$, respectively.

Now a coordinate transformation from the retarded time \(u\) to the
usual time coordinate \(t=u+r\) converts the above metric
into the NGT static spherically symmetric metric \cite{Moff91,Moff79-2}:
\begin{eqnarray} \label{NGTSchwarz}
{ds}^2 &=& (1 + \frac{l_{s}^4}{r^4})(1 - \frac{2M_{s}}{r}){dt}^2 - {(1 -
\frac{2M_{s}}{r})}^{-1}{dr}^2 \nonumber \\
&& - r^2 \left( {d\theta}^2 + {\sin}^2\theta{d\phi}^2 \right),
\end{eqnarray}
and
\[
 g_{[01]} = \frac{{l_{s}^2}}{{r^2}}.
\]
Thus, the static spherically symmetric limit, \(M_{s}\), of the
``mass aspect'' \(M(u,\theta)\) can only be interpreted as the mass
of the system. Similarly, the static spherically symmetric limit,
$l_{s}^{2}$, of the ``charge aspect'' $W_{2}(u,\theta)$ is identically
the NGT charge of the system as shown by equations (\ref{ngtc}) and
(\ref{ngtcc}).

If we define the mass \(m(u)\) of the system as the mean value of
\(M(u,\theta)\) over the sphere:
\begin{equation} \label{mass}
m(u)=\frac{1}{2} \int^{\pi}_{0} M(u,\theta)\sin\theta d\theta =
<M(u,\theta)> ,
\end{equation}
then \(c(u,\theta)\) completely determines the time evolution of
the mass \(m(u)\). Integrating (\ref{massderiv}) and noticing that
the second term does not contribute to the integral due to
the condition that $c$ be regular on the polar axis, we get
\begin{equation} \label{centralresult}
m_{,u}=\frac{dm}{du}=-\frac{1}{2} \int^{\pi}_{0} {c_{,u}}^{2}
\sin\theta d\theta .
\end{equation}
Since we discussed here systems whose initial and final states are
static, the physical interpretation of \(m(u)\) as the mass of the
system is unambiguous. Analogously to the GR case the main result
is as follows:

{\em The mass of an axially symmetric NGT system is constant only
if the system remains static. If the system evolves in time (emits
waves), the mass decreases monotonically.}

Since radiation is the only energy loss mechanism available to the
system, the above proves that gravitational waves emitted by an
axi--symmetric reflexion symmetric NGT source compatible with the
metric (\ref{genmetric}) carry positive energy or, in other words,
the flux of gravitational energy in NGT is positive.

\section*{Conclusions} \label{con}

We have proved that an NGT axi--symmetric system emitting
gravitational waves has the usual GR-like asymptotic behaviour in
the wave zone. The NGT contributions to the physical quantities
decay rapidly with the distance from the source and the energy flux
at spatial infinity is necessarily positive.

While we concentrated on an axi-symmetric source to simplify computations,
the validity of our result is not confined to this particular symmetry.
This contention can be made concrete by adapting the analysis made by
Sachs \cite{Sachs}, where it was shown that the axi-symmetric solution
contains all the important features of any isolated, radiative system in GR.
Sachs' result follows from considerations of the asymptotic nature of the
spacetime, and these considerations are unchanged in NGT.
Physically, we can argue that an axi-symmetric source provides a complete
range of multipole moments to act as a source of gravitational waves, and
thus provides a general test of the wave sector of any gravitational theory.
Moreover, in the wave zone, the superposition principle may be used to
construct the radiation pattern of any isolated body from a suitable sum
of axi-symmetric solutions.
Our result totally refutes the recent claims that NGT has bad wave behaviour
\cite{DDM1,Dam}, and shows that aesthetically unappealing,
phenomenological modifications to NGT \cite{DDM2} are not necessary.

\section*{Acknowledgements}
This work was supported by the Natural Sciences and Engineering
Research Council of Canada. One of the authors (NJC) is
grateful for the support provided by a Canadian Commonwealth
Scholarship. We thank P. Savaria for helpful discussions.

\appendix
\widetext
\section*{}

\begin{eqnarray}
\hspace*{-2cm}\Gamma^{0}_{00}=& & 2{B_{1}}_{,u}r^{-1}-\left(U_{1}U_{2}+
{U_{1}}^2c+ M-2{B_{2}}_{,u}-B_{1} \right)r^{-2} \nonumber \\
&+& \left[ 2B_{2}-2MB_{1}+V_{1}-2{L_{0}}^{2}-{U_{2}}^{2}
+2{B_{3}}_{,\theta}\right. \nonumber \\ & & \left. +2U_{1}\left(
U_{2}B_{1}-U_{3}-2cU_{2}\right)+2c{U_{1}}^{2}\left(B_{1}-c\right)
\right]r^{-3} + ...,\\
\Gamma^{0}_{(01)}=&-& \left( S_{2}L_{0}+2W_{2}^2 \right)r^{-5} + ...,\\
\Gamma^{0}_{[01]}=&-& 2W_{2}r^{-3} - \left( 3W_{3} +\frac{1}{2}
S_{2}U_{1} -6B_{1}W_{2} \right)r^{-4}+ ...,\\
\Gamma^{0}_{(02)}=&-& \frac{1}{2}U_{1} + \left( B_{1}U_{1} +
{B_{1}}_{,\theta} \right)r^{-1} \nonumber \\& &+ \left[ cU_{2}+
{B_{2}}_{,\theta} +\frac{1}{2}U_{3}+U_{1}\left(B_{2}-{B_{1}}^2+c^2\right)
\right]r^{-2} + ..., \\
\Gamma^{0}_{[02]}=&-& L_{0}r^{-1} + \left[L_{0}(c+2 B_{1}) +
\frac{1}{2}\left({W_{2}}_{,\theta}+{S_{2}}_{,u}+U_{1}W_{2}-3L_{1}
\right)\right]r^{-2}+ ...,\\
\Gamma^{0}_{(12)}=& & 2W_{2}S_{2}r^{-5}+...,\\
\Gamma^{0}_{[12]}=&-& 3S_{2}r^{-3} + S_{2}\left(8B_{1}+c\right)r^{-4}..., \\
\Gamma^{0}_{22}=& &c +r - 2B_{1} -2\left(B_{2}+cB_{1}-{B_{1}}^{2}\right)
r^{-1} \nonumber \\ &+&\left[2c{B_{1}}^{2}+4B_{2}B_{1}-2B_{2}c-C
-\frac{1}{2}c^{3}-2B_{3}-\frac{4}{3}{B_{1}}^{3}\right]r^{-2} + ..., \\
\Gamma^{0}_{33}/\sin^{2}\theta=& & r - c - 2B_{1} +2\left( cB_{1}-B_{2}
+{B_{1}}^{2}\right)r^{-1} \nonumber \\
&+& \left[4B_{2}B_{1}-2c{B_{1}}^{2}+2B_{2}c+C+\frac{1}{2}c^{3}
-2B_{3}-\frac{4}{3}{B_{1}}^{3}\right]r^{-2}  + ..., \\
\Gamma^{1}_{00}=& & \left[ {U_{1}}^2\left({U_{1}}_{,\theta}+c_{,u}
\right)-{B_{1}}_{,u}-M_{,u}\right]r^{-1}  \nonumber
\\&+& \left[ {U_{1}}^2 \left(2(c-B_{1})(c_{,u}+{U_{1}}_{,\theta})+
{U_{2}}_{,\theta}+c+c_{,\theta}U_{1} \right) +  \frac{1}{2}{V_{1}}_{,u}
-{B_{2}}_{,u}-B_{1} \right. \nonumber  \\& & \left. +M(1+2{B_{1}}_{,u})
+U_{1}U_{2}\left(1+2c_{,u}+2{U_{1}}_{,\theta}\right)+U_{1}
{(M-B_{1})}_{,\theta} \right]r^{-2} + ..., \nonumber \\ & & \\
\Gamma^{1}_{(01)}=& & \frac{1}{2}{U_{1}}^2r^{-1} + \left[{U_{1}}^2
(c-B_{1})+M-B_{1}+U_{1}\left(\frac{3}{2}U_{2}-{B_{1}}_{,\theta}\right)
\right]r^{-2} \nonumber \\
&+& \left[{U_{1}}^{2}\left(c^{2}+{B_{1}}^{2}-B_{2}-2cB_{1}\right)
+3U_{1}U_{2}\left(c-B_{1} \right)-V_{1}-2B_{2} \right. \nonumber \\ & &
\left. +2MB_{1}+{U_{2}}^{2}
-U_{1}{B_{2}}_{,\theta}-U_{2}{B_{1}}_{,\theta}+2U_{1}U{3}+{L_{0}}^{2}
\right]r^{-3} + ..., \\
\Gamma^{1}_{[01]}=& & \left({W_{2}}_{,u} -\frac{1}{2}L_{0}U_{1}\right)r^{-2}
\nonumber \\&+& \left[W_{2}\left(2-2{B_{1}}_{,u}+\frac{1}{2}{U_{1}}^2\right)
-L_{0}\left(U_{2}+{B_{1}}_{,\theta}\right) - 2B_{1}{W_{2}}_{,u} \right.
\nonumber \\& & \left.+U_{1} \left(\frac{1}{2}{W_{2}}_{,\theta}-\frac{1}{2}
{S_{2}}_{,u}-L_{1}+L_{0}B_{1} \right) +{W_{3}}_{,u}-S_{2}{U_{1}}_{,u}\right]
r^{-3}+ ..., \\
\Gamma^{1}_{(02)}=& & \frac{1}{2}U_{1} \left(1-2{U_{1}}_{,\theta}-
2c_{,u}\right) \nonumber \\&+& \left[U_{1}\left(B_{1}\left(2c_{,u}-1\right)
+2{U_{1}}_{,\theta}(B_{1}-c)-{U_{2}}_{,\theta}-M-2cc_{,u}\right) \right.
\nonumber \\& & \left.-U_{2} \left(c_{,u}+{U_{1}}_{,\theta}\right)-M_{,\theta}
-{U_{1}}^2c_{,\theta} \right]r^{-1} \nonumber \\
&+& \left[ W_{2}{L_{0}}_{,u}-L_{0}
{L_{0}}_{,\theta} +\frac{1}{2}{V_{1}}_{,\theta}
-U_{3}\left(\frac{1}{2}+{U_{1}}_{,\theta}+c_{,u}\right) \right. \nonumber \\
& &+U_{2}\left(2B_{1}c_{,u}-2cc_{,u}+2B_{1}{U_{1}}_{,\theta}
-2c{U_{1}}_{,\theta}-{U_{2}}_{,\theta}-c \right) \nonumber \\
& & +U_{1}\left( 2MB_{1}-2U_{2}c_{,\theta}-\frac{3}{2}c^{2}c_{,u}-c^{2}
-B_{2}+{B_{1}}^{2}-C_{,u}-{U_{3}}_{,\theta} \right. \nonumber \\
& & +\frac{1}{2}V_{1}-\frac{3}{2}{L_{0}}^{2}
+2\left(B_{1}-c\right)\left({U_{2}}_{,\theta}+U_{1}c_{,\theta}\right)
-2c^{2}{U_{1}}_{,\theta} \nonumber \\ & & \left. \left.
+2\left(2cB_{1}+B_{2}-{B_{1}}^{2}\right)
\left(c_{,u}+{U_{1}}_{,\theta}\right)\right)\right] r^{-2} + ..., \\
\Gamma^{1}_{[02]}=& & {L_{0}}_{,u} + \left[{L_{1}}_{,u} + L_{0}\left(
1-c_{,u}-2{B_{1}}_{,u}\right)+{L_{0}}_{,\theta}U_{1}-2B_{1}{L_{0}}_{,u}
\right]r^{-1} + ..., \\
\Gamma^{1}_{11}=& & -2B_{1}r^{-2}-4B_{2}r^{-3}-6B_{3}r^{-4}
+2\left(S_{2}L_{0}+2{W_{2}}^{2}\right)r^{-5} + ...,\\
\Gamma^{1}_{(12)}=&-& \frac{1}{2}U_{1}+\left[{B_{1}}_{,\theta}-U_{2}+
U_{1}(B_{1}-c)\right]r^{-1} \nonumber \\
&+&\left[ U_{1}\left(B_{2}-{B_{1}}^{2}-c^{2}+2cB_{1}\right)
+2U_{2}\left(B_{1}-c\right)-\frac{3}{2}U_{3}+{B_{2}}_{,\theta}
\right]r^{-2} + ..., \\
\Gamma^{1}_{[12]}=& & \frac{1}{2}\left(U_{1}W_{2}-L_{1}-{W_{2}}_{,\theta}
+{S_{2}}_{,u}\right)r^{-2} + ..., \\
\Gamma^{1}_{22}=&-& r(1-c_{,u}-{U_{1}}_{,\theta}) \nonumber \\
 &+& \left[2M +2c_{,u}(c-B_{1})
+2B_{1}(1-{U_{1}}_{,\theta}) -c(1-2{U_{1}}_{,\theta})+
{U_{2}}_{,\theta} +U_{1}c_{,\theta}\right] \nonumber \\
&+& \left[ 2B_{2}-2{B_{1}}^{2}+2cB_{1}-4MB_{1}-V_{1}+2{L_{0}}^{2}
+2Mc +C_{,u} \right. \nonumber \\ & &+c_{,u}\left(2{B_{1}}^{2}-2B_{2}-4cB_{1}
+\frac{3}{2}c^{2}\right)
+ c_{,\theta}\left(U_{2}+2cU_{1}-2U_{1}B_{1}\right) \nonumber \\
& & \left. +2{U_{1}}_{,\theta} \left(c^{2}+{B_{1}}^{2}-2cB_{1}-B_{2}\right)
+2{U_{2}}_{,\theta}\left(c-B_{1}\right) +{U_{3}}_{,\theta}
\right] r^{-1} + ..., \\
\Gamma^{1}_{33}/\sin^{2}\theta=&-& r(1+c_{,u}-U_{1}\cot\theta)
\nonumber \\ &+& \left[ 2M + c
-c_{,\theta}U_{1}+ 2c_{,u}(c+B_{1}) + 2B_{1} - \left(2U_{1}
(c+B_{1})-U_{2} \right)\cot\theta \right] \nonumber \\
&+& \left[ 2B_{2}-2{B_{1}}^{2}-2cB_{1}-4MB_{1}-V_{1}
-2Mc-C_{,u} \right. \nonumber \\ & &-c_{,u}\left(2{B_{1}}^{2}-2B_{2}+4cB_{1}
+\frac{3}{2}c^{2}\right)
-c_{,\theta}\left(U_{2}-2cU_{1}-2U_{1}B_{1}\right) \nonumber \\
& & \left. +\left(2U_{1} \left(c^{2}+{B_{1}}^{2}+2cB_{1}-B_{2}\right)
-2U_{2}\left(c+B_{1}\right) +U_{3}\right)\cot\theta
\right] r^{-1} + ..., \\
\Gamma^{2}_{00}=&-& {U_{1}}_{,u}r^{-1} +\left[U_{1}\left( 2{B_{1}}_{,u}
-2c_{,u}-{U_{1}}_{,\theta}\right)-{U_{2}}_{,u}\right]r^{-2} \nonumber \\
&+& \left[ {B_{1}}_{,\theta}-M_{,\theta}-{U_{3}}_{,u}-{U_{1}}^{2}
\left(U_{2}+cU_{1}-c_{,\theta}\right)+U_{2}\left(2{B_{1}}_{,u}-2c_{,u}
-{U_{1}}_{,\theta}\right) \right. \nonumber \\
& & \left. +U_{1}\left(B_{1}-{U_{2}}_{,\theta}+2{B_{2}}_{,u}\right) \right]
r^{-3}  + ..., \\
\Gamma^{2}_{(01)}=&-& \frac{1}{2}U_{1}r^{-2} + \left({B_{1}}_{,\theta}+
U_{1}c\right)r^{-3}\nonumber \\ &+& \left[ cU_{2}+\frac{1}{2}U_{3}
+{B_{2}}_{,\theta}+2{B_{1}}_{,\theta}\left(B_{1}-c\right) \right]r^{-4}+ ... ,
\\
\Gamma^{2}_{[01]}=& & L_{0}r^{-3}-\left[3(U_{1}W_{2}+cL_{0})+\frac{1}{2}
\left({W_{2}}_{,\theta}-{S_{2}}_{,u}-3L_{1}\right)\right]r^{-4} + ..., \\
\Gamma^{2}_{(02)}=& & \left(c_{,u}-\frac{1}{2}{U_{1}}^2\right)r^{-1}
+U_{1}\left(B_{1}U_{1}+{B_{1}}_{,\theta}-\frac{1}{2}U_{2}\right)r^{-2}
\nonumber \\
&+& \left[ C_{,u}-\frac{1}{2}c^{2}c_{,u}+{L_{0}}^{2}+U_{1}{B_{2}}_{,\theta}
+U_{2}{B_{1}}_{,\theta}\right. \nonumber \\
& & \left. +{U_{1}}^{2}\left(B_{2}+c^{2}-{B_{1}}^{2}\right)
+U_{1}U_{2}\left(B_{1}+c\right) \right]r^{-3}
+ ..., \\
\Gamma^{2}_{[02]}=&-&  \left({L_{0}}_{,\theta}+\frac{1}{2}U_{1}L_{0}
\right)r^{-2} + ..., \\
\Gamma^{2}_{(11)}=& &4W_{2}S_{2} r^{-7} + ..., \\
\Gamma^{2}_{(12)}=& & r^{-1} -cr^{-2} +\left(\frac{1}{2}c^{3}-3C\right)
r^{-4} + ..., \\
\Gamma^{2}_{[12]}=&-& W_{2}r^{-3}+\left(2W_{2}B_{1}-W_{3}+cW_{2}
-\frac{5}{2}S_{2}U_{1}-{S_{2}}_{,\theta}\right)r^{-4} + ..., \\
\Gamma^{2}_{22}=& & U_{1} + \left[U_{1}(c-2B_{1})+U_{2}+c_{,\theta}\right]
r^{-1} \nonumber \\
&+& \left[ U_{3}+U_{2}\left(c-2B_{1}\right)+2U_{1}\left({B_{1}}^{2}
-B_{2}-cB_{1}\right)\right]r^{-2} + ..., \\
\Gamma^{2}_{33}/\sin^2\theta=& & U_{1} - \cot\theta  +\left(U_{2}+c_{,\theta}
-cU_{1}-2B_{1}U_{1}+4c\cot\theta\right)r^{-1} \nonumber \\
&+& \left[ U_{3}-U_{2}\left(c+2B_{1}\right)+2U_{1}\left({B_{1}}^{2}
-B_{2}+cB_{1}\right)\right. \nonumber \\
& & \left. -4c\left(c_{,\theta}+2c\cot\theta\right)\right]r^{-2}
 + ..., \\
\Gamma^{3}_{(03)}=&-&c_{,u}r^{-1} +\left(\frac{1}{2}c^2c_{,u}-C_{,u}\right)
r^{-3} + ..., \\
\Gamma^{3}_{[03]}=& & L_{0}(U_{1}-\cot\theta)r^{-2} \nonumber \\&+& \left[
W_{2}(U_{1}\cot\theta-1-c_{,u})+ L_{1}(U_{1}-\cot\theta) \right. \nonumber \\
& & \left.+L_{0}(c_{,\theta}-2B_{1}U_{1}+2c\cot\theta +cU_{1}+U_{2})
\right]r^{-3} + ..., \\
\Gamma^{3}_{(13)}=& & r^{-1} + cr^{-2} + \left(3C-\frac{1}{2}c^{3}\right)
r^{-4} + ..., \\
\Gamma^{3}_{[13]}=&-& W_{2}r^{-3}+ \left[W_{2}\left(2B_{1}-c\right)
+S_{2}\left(U_{1}-\cot\theta\right) -W_{3}\right]r^{-4} + ..., \\
\Gamma^{3}_{(23)}=& & \cot\theta - c_{,\theta}r^{-1} +
\left(\frac{1}{2}c^{2}c_{,\theta}-C_{,\theta}\right)r^{-3} + ..., \\
\Gamma^{3}_{[23]}=&-& L_{0}r^{-1} +\left[L_{0}(2B_{1}-c)-L_{1}\right]r^{-2}
+ ...,
\end{eqnarray}

\narrowtext


\begin{references}
\bibitem{endnote} The full static spherically symmetric solution in NGT
admits a term $g_{[23]}\sim r^2 \sin\theta$, which is not compatible with
asymptotic inhomogeneous orthochronous Lorentz invariance.
It is important to impose such
an invariance on the wave solutions so as to facilitate a definition of
energy. One may keep the three functions $g_{[03]},\; g_{[13]},
\; g_{[23]}$ in the calculation and maintain asymptotic invariance, if:
$g_{[03]}=D_{0}+D_{1}/r+\dots$, $g_{[13]}=E_{0}+E_{1}/r+\dots$, and
$g_{[23]}=r F_{-1}+F_{0}+\dots$, where $D_{i},\; E_{i},\; F_{i}$ are
functions of $u$ and $\theta$. Repeating the calculation in ref. 2 with
all six skew functions gives the following: At lowest order the skew
divergence equation, $(\sqrt{-g}g^{[3 \nu]})_{,\nu}=0$, gives
$E_{0,\, u}=0 \Rightarrow E_{0}=0$. The lowest
order in $R_{\{[02],3]\}}=0$ gives $F_{-1\, ,uu}=0 \Rightarrow F_{-1}=0$ (as
demanded by the quasi-periodic boundary conditions). The next order in
the skew divergence equations then gives $E_{1\, ,u}=0 \Rightarrow E_{1}=0$.
Putting all this into $R_{\{[01],3\}}=0$ yields $D_{0\, ,u}=0
\Rightarrow D_{0}=0$. Since the two sets of three skew functions have
remained uncoupled to this order, the solutions for $g_{[01]},\; g_{[02]},
\; g_{[12]}$ are unchanged from those in ref. 2.
At this stage our work is done since all leading
terms ($1/r$ terms in cartesian coordinates) have vanished. It is thus a
trivial exercise to show that $g_{[03]},\; g_{[13]}, \; g_{[23]}$
can play no role in determining the flux of gravitational radiation.
\end{references}
\end{document}